\documentclass[12pt]{article}
\usepackage{amsmath,amssymb}
\usepackage{epsfig} 

\pdfoutput=1
\usepackage{graphicx, rotating}
\usepackage{hyperref}
\usepackage{slashed}
\usepackage{ifpdf}

\ifx\pdfoutput\undefined
\usepackage[dvips,bookmarks=false]{hyperref}	
\else
\usepackage{hyperref}	
\fi
\hypersetup{colorlinks,bookmarksopen,bookmarksnumbered,citecolor=verdes,
linkcolor=blus,pdfstartview=FitH,urlcolor=rossos}
\def\hhref#1{\href{http://arxiv.org/abs/#1}{#1}} 

\usepackage{amsmath}
\usepackage{slashed}

\usepackage{cite}

\newcommand{\beq}{\begin{equation}}
\newcommand{\eeq}{\end{equation}}
\newcommand{\fig}[1]{~\ref{fig:#1}}

\newcount\Mac  \Mac=1  
\newcommand{\ifMac}[2]{\ifnum\Mac=1 #1 \else #2 \fi}
\def\putps(#1,#2)(#3,#4)#5#6{\ifnum\Mac=1 \put(#1,#2){\special{picture #5}}
\else  \put(#3,#4){\includegraphics{#6}} \fi}

\newcommand{\One}{\hbox{1\kern-.24em I}}

\newcommand{\GeV}{\,{\rm GeV}}

\newcommand{\eV}{\,{\rm eV}}

\newcommand{\eq}[1]{~{\rm(\ref{eq:#1})}}

\newcommand{\lascia}[1]{}
\makeatletter
\def\art{\@ifnextchar[{\eart}{\oart}}
\def\eart[#1]#2#3#4#5#6{{\rm #2}, {#3 #4} {\rm (#6) #5} [arXiv:{\hhref{#1}}]}
\def\hepart[#1]#2{{\rm #2, arXiv:\hhref{#1}}}
\newcommand{\oart}[5]{{\rm #1}, {#2 #3} {\rm (#5) #4}}

\newcounter{alphaequation}[equation]
\def\thealphaequation{\theequation\hbox to
0.6em{\hfil\alph{alphaequation}\hfil}}
\def\eqnsystem#1{
\def\@eqnnum{{\rm (\thealphaequation)}}
\def\@@eqncr{\let\@tempa\relax \ifcase\@eqcnt \def\@tempa{& & &} \or
  \def\@tempa{& &}\or \def\@tempa{&}\fi\@tempa
  \if@eqnsw\@eqnnum\refstepcounter{alphaequation}\fi
\global\@eqnswtrue\global\@eqcnt=0\cr}
\refstepcounter{equation} \let\@currentlabel\theequation \def\@tempb{#1}
\ifx\@tempb\empty\else\label{#1}\fi
\refstepcounter{alphaequation}
\let\@currentlabel\thealphaequation
\global\@eqnswtrue\global\@eqcnt=0 \tabskip\@centering\let\\=\@eqncr
$$\halign to \displaywidth\bgroup \@eqnsel\hskip\@centering
$\displaystyle\tabskip\z@{##}$&\global\@eqcnt\@ne
\hskip2\arraycolsep\hfil${##}$\hfil& \global\@eqcnt\tw@\hskip2\arraycolsep
$\displaystyle\tabskip\z@{##}$\hfil
\tabskip\@centering&\llap{##}\tabskip\z@\cr}

\def\endeqnsystem{\@@eqncr\egroup$$\global\@ignoretrue} \makeatother

\def\Lag{{\cal L}}

\def\Tr{\mathop{\rm Tr}}
\def\circa#1{\,\raise.3ex\hbox{$#1$\kern-.75em\lower1ex\hbox{$\sim$}}\,}

\usepackage{multicol}
\usepackage{color}
\definecolor{rosso}{cmyk}{0,1,1,0.4}
\definecolor{rossos}{cmyk}{0,1,1,0.55}
\definecolor{rossoc}{cmyk}{0,1,1,0.2}
\definecolor{blu}{cmyk}{1,1,0,0.3}
\definecolor{blus}{cmyk}{1,1,0,0.6}
\definecolor{bluc}{cmyk}{1,1,0,0.1}
\definecolor{verde}{cmyk}{0.92,0,0.59,0.25}
\definecolor{verdec}{cmyk}{0.92,0,0.59,0.15}
\definecolor{verdes}{cmyk}{0.92,0,0.59,0.4}
\definecolor{grigio}{cmyk}{0,0,0,0.07}
\definecolor{rosa}{cmyk}{0,0.1,0.1,0.02}
\definecolor{rosino}{cmyk}{0,0.05,0.05,0.02}
\definecolor{rosas}{cmyk}{0,0.3,0.25,0.05}
\definecolor{celeste}{cmyk}{0.1,0,0,0.02}
\definecolor{giallino}{cmyk}{0,0,0.4,0.02}
\definecolor{rosso}{cmyk}{0,1,1,0.4}
\definecolor{rossos}{cmyk}{0,1,1,0.55}
\definecolor{rossoc}{cmyk}{0,1,1,0.2}
\definecolor{blu}{cmyk}{1,1,0,0.3}
\definecolor{bluc}{cmyk}{1,1,0,0.1}
\definecolor{blucc}{cmyk}{0.7,0.5,0,0}
\definecolor{viola}{cmyk}{0,1,0,0.6}
\definecolor{viola2}{cmyk}{0,1,0.2,0.6}
\definecolor{verde}{cmyk}{0.92,0,0.59,0.25}
\definecolor{verdec}{cmyk}{0.92,0,0.59,0.15}
\definecolor{verdes}{cmyk}{0.92,0,0.59,0.4}
\definecolor{verdino}{cmyk}{0.12,0,0.09,0.05}
\definecolor{giallo}{cmyk}{0,0,1,0}
\definecolor{gialloverde}{cmyk}{0.44,0,0.74,0}

\font\tenrsfs=rsfs10 at 12pt

\font\sevenrsfs=rsfs7
\font\fiversfs=rsfs5
\newfam\rsfsfam
\textfont\rsfsfam=\tenrsfs
\scriptfont\rsfsfam=\sevenrsfs
\scriptscriptfont\rsfsfam=\fiversfs
\def\mathscr#1{{\fam\rsfsfam\relax#1}}
\def\Lag{\mathscr{L}}

\def\eq#1{eq.~(\ref{#1})}
\def\beq{\begin{equation}}
\def\eeq{\end{equation}}
\def\bea{\begin{eqnarray}}
\def\eea{\end{eqnarray}}

\def\m@th{\mathsurround=0pt }
\def\leftrightarrowfill{$\m@th \mathord\leftarrow \mkern-6mu
        \cleaders\hbox{$\mkern-2mu \mathord- \mkern-2mu$}\hfill
        \mkern-6mu \mathord\rightarrow$}

\def\overleftrightarrow#1{\vbox{\ialign{##\crcr
        \leftrightarrowfill\crcr\noalign{\kern-1pt\nointerlineskip}
        $\hfil\displaystyle{#1}\hfil$\crcr}}}

\def\simgt{\stackrel{>}{{}_\sim}}

\newcommand{\be}{\begin{equation}}
\newcommand{\ee}{\end{equation}}

\def\shat{\ifmmode \hat{s}\else $\hat{s}$\fi}
\def\gp2{{g'}^2}
\def\g2{g^2}
\def\g32{g_s^2}

\newcommand{\newc}{\newcommand}

\newc{\gsim}{\lower.7ex\hbox{$\;\stackrel{\textstyle>}{\sim}\;$}}
\newc{\lsim}{\lower.7ex\hbox{$\;\stackrel{\textstyle<}{\sim}\;$}}
\newc{\ie}{{\it i.e.}}
\newc{\etal}{{\it et al.}}
\newc{\mev}{\hbox{\rm\,MeV}}
\newc{\gev}{\hbox{\rm\,GeV}}
\newc{\tev}{\hbox{\rm\,TeV}}
\newc{\xpb}{\hbox{\rm\, pb}}
\newc{\xfb}{\hbox{\rm\, fb}}

\newc{\G}{{\cal G}}
\newc{\h}{{\cal H}}
\newc{\D}{{\cal D}}
\newc{\E}{{\cal E}}

\newc{\mtop}{m_t}
\newc{\mbot}{m_b}
\newc{\mz}{M_Z}
\newc{\mw}{M_W}
\newc{\alphasmz}{\alpha_s(M_Z)}
\newc{\swsq}{\sin^2\theta_W}
\newc{\cwsq}{\cos^2\theta_W}
\newc{\tw}{\tan\theta_W}
\newc{\cw}{\cos\theta_W}
\newc{\sw}{\sin\theta_W}
\newc{\BR}{\hbox{\rm BR}}
\newc{\zbb}{Z\to b\bar}
\newc{\Gb}{\Gamma (Z\to b\bar b)}
\newc{\Gh}{\Gamma (Z\to \hbox{\rm hadrons})}
\newc{\sgn}{\mbox{sgn}}

\def\eq#1{eq.~(\ref{#1})}

\newcounter{mysubequation}[equation]

\def\beq{\begin{equation}}
\def\eeq{\end{equation}}
\def\bea{\begin{eqnarray}}
\def\eea{\end{eqnarray}}
\def\slashchar#1{\setbox0=\hbox{$#1$}           
   \dimen0=\wd0                                 
   \setbox1=\hbox{/} \dimen1=\wd1               
   \ifdim\dimen0>\dimen1                        
      \rlap{\hbox to \dimen0{\hfil/\hfil}}      
      #1                                        
   \else                                        
      \rlap{\hbox to \dimen1{\hfil$#1$\hfil}}   
      /                                         
   \fi}                                         %
\catcode`@=11
\long\def\@caption#1[#2]#3{\par\addcontentsline{\csname
  ext@#1\endcsname}{#1}{\protect\numberline{\csname
  the#1\endcsname}{\ignorespaces #2}}\begingroup
    \small
    \@parboxrestore
    \@makecaption{\csname fnum@#1\endcsname}{\ignorespaces #3}\par
  \endgroup}
\catcode`@=12

\begin{document}

\color{black}
\vspace{1cm}
\begin{center}
{\Huge\bf\color{rossos} Higgs mass implications on the stability of the electroweak vacuum}\\
\bigskip\color{black}\vspace{0.6cm}{
{\large\bf  Joan Elias-Mir\'o$^a$, Jos\'e R. Espinosa$^{a,b}$, Gian F. Giudice$^{c}$,\\ [2mm]  Gino Isidori$^{c,d}$, Antonio Riotto$^{c,e}$, Alessandro Strumia$^{f,g}$}
\vspace{0.5cm}
} \\[7mm]
{\em $(a)$ {IFAE, Universitat Aut\'onoma de Barcelona, 08193 Bellaterra, Barcelona, Spain}}\\
{\em $(b)$ {ICREA, Instituci\`o Catalana de Recerca i Estudis Avan\c{c}ats, Barcelona, Spain}}\\
{\em $(c)$ {CERN, Theory Division, CH--1211 Geneva 23,  Switzerland}}\\
{\it $(d)$ INFN, Laboratori Nazionali di Frascati, Via E.~Fermi 40, Frascati, Italy}\\
{\em $(e)${INFN, Sezione di Padova, Via Marzolo 8, I-35131 Padua, Italy}}\\
{\it $(f)$ Dipartimento di Fisica dell'Universit{\`a} di Pisa and INFN, Italy}\\
{\it  $(g)$ National Institute of Chemical Physics and Biophysics, Ravala 10, Tallinn, Estonia}\\
\end{center}
\bigskip
\centerline{\large\bf\color{blus} Abstract}
\begin{quote}\small
We update instability and metastability bounds of the Standard Model electroweak vacuum in 
view of the recent ATLAS and CMS Higgs  results. For a Higgs mass in the range 
124--126 GeV, and for the current central values of the top mass and strong coupling constant,  
the Higgs potential develops an instability around $10^{11}$ GeV, with a lifetime much longer than the age of the Universe.
However, taking into account theoretical and
experimental errors, stability up to the  Planck scale cannot be excluded.
Stability at finite temperature implies an  upper  bound on the reheat temperature after inflation, which depends critically on the precise values of the Higgs and top masses. 
A Higgs  mass in the range 124--126 GeV
 is compatible with very high values of the reheating temperature, without
conflict with mechanisms of baryogenesis such as leptogenesis.
We derive an upper bound on the mass of heavy right-handed neutrinos
by requiring that  their Yukawa couplings do not destabilize the Higgs potential.
\end{quote}

\normalsize



\section{Introduction}

Experimental data recently reported by the LHC experiments after the analysis of their 5/fb dataset
restrict the Standard Model (SM) Higgs boson mass to be in the range
$115~\gev < m_h < 131~\gev$~(ATLAS  \cite{ATLAS}) and
$m_h < 127~\gev$~(CMS \cite{CMS}), with a first  
hint in the mass window $124~{\rm GeV} < m_h <126$~GeV. 
Such a light Higgs is in good agreement with the indirect indications 
derived from electroweak precisions constraints~\cite{EW} under the hypothesis of negligible 
contributions of physics beyond the SM. Moreover, no clear signal of non-SM physics
has emerged yet from collider searches.

Motivated by this experimental situation, we present here a detailed investigation 
about the stability of the Standard Model vacuum under the hypothesis 
$124~{\rm GeV} < m_h <126$~GeV, assuming  
the validity of the SM up to very high energy scales.

Despite the fact that there is no evidence for physics beyond the Standard Model (SM) from the LHC, the experimental information 
on the Higgs mass gives us useful hints on the structure of the theory at very short distances,
%
%
thanks to the sizable logarithmic variation  of the Higgs quartic coupling at high energies.
For instance, the recent LHC results can be used to constrain the scale of supersymmetry breaking,
even when this scale is far beyond the TeV range~\cite{GS}.
Another consideration is that, for low enough values of $m_h$,  
the Higgs potential can develop an instability at high field values,
signaling an unambiguous inconsistency of the model at very short distances~\cite{con,thermal1,AI,CEQ,Hambye,IRS,IRST,Jones,noi,latbound}. 

In this paper, we reconsider the instability and metastability bounds.
Then we study the implication of these results on
information about the early
stages of the Universe. 
If the Universe spent a period of its evolution in the presence of a hot thermal plasma, the absence of excessive
thermal Higgs field fluctuations, which might destabilize our present vacuum, imposes an upper  bound on the
reheat temperature after inflation, generically denoted by  $T_{\rm RH}$. This upper bound has implications
for the dynamics of the evolution of the Universe and for the creation of the observed baryon asymmetry. In particular, these considerations become relevant in the case of leptogenesis (for a review see Ref.~\cite{lepto,nureview}), in which the decays of heavy right-handed (RH)  neutrinos  are responsible for the generation of a lepton asymmetry, eventually leading to a  baryon asymmetry. Depending on the mass of the RH neutrinos, the leptogenesis scenario might require relatively high values of the
reheating temperature. 

The presence of massive RH neutrinos has also a direct impact on the structure of the Higgs potential at high energies, 
and thus on the stability bounds, via one-loop corrections induced by the RH neutrino Yukawa couplings.
Independently on any consideration of leptogenesis or reheating temperature, the requirement that the electroweak vacuum has a lifetime longer than the age of the Universe implies an interesting upper bound on the mass of the RH neutrinos, as a function of the physical neutrino mass. We derive this limit assuming that the SM and the RH neutrinos describe all the degrees of freedom up to a very large energy scale, close to the Planck mass.

\begin{figure}[t]
$$\includegraphics[width=0.5\textwidth]{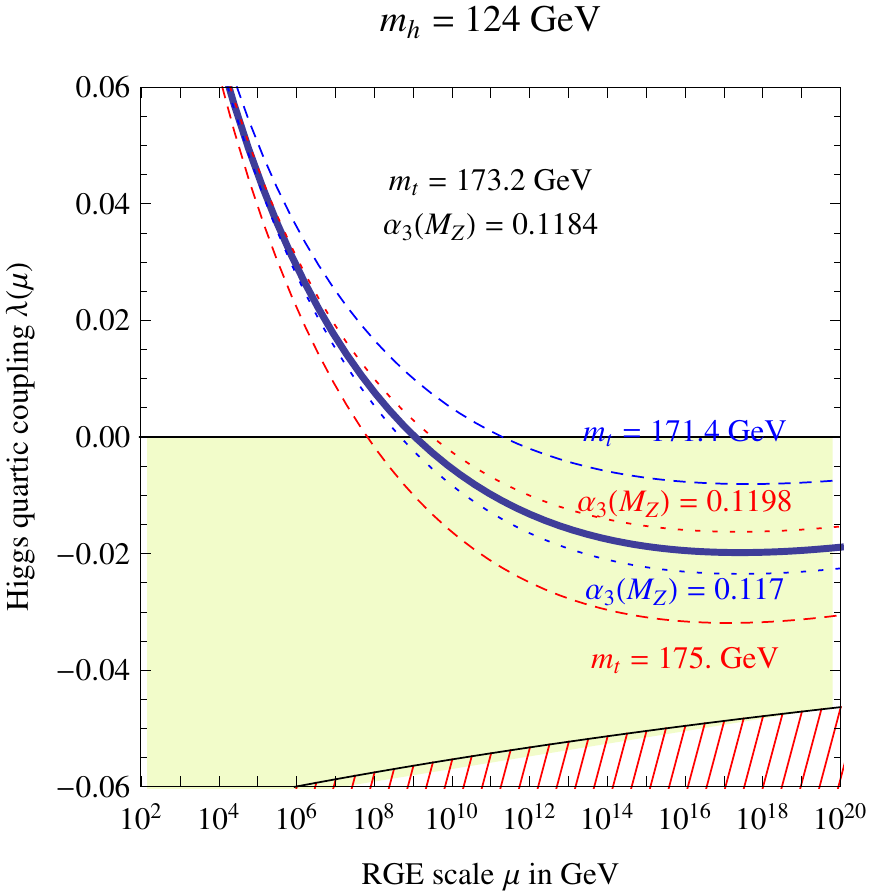}\qquad \includegraphics[width=0.5\textwidth]{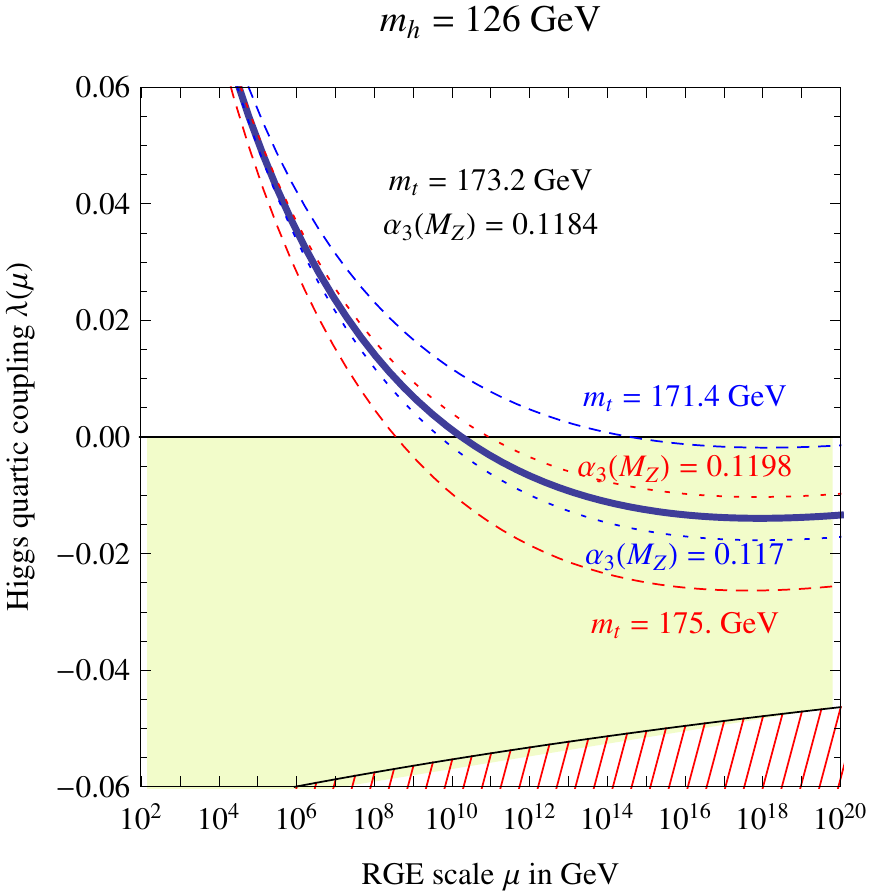}$$
\caption{\em RG evolution of the Higgs self coupling, for different Higgs masses
for the central value of $m_t$ and $\alpha_s$, as well as for
$\pm 2\sigma$ variations of $m_t$ (dashed lines) and  $\alpha_s$ (dotted lines).
For negative values of $\lambda$, 
the life-time of the SM vacuum due to quantum tunneling at zero temperature
is longer than the age of the Universe as long as $\lambda$ remains 
above the region shaded in red, which takes into account the finite corrections to the 
effective bounce action renormalised at the same scale as $\lambda$  (see~\cite{IRS} 
for more details). 
\label{fig:run} }
\end{figure}

\begin{figure}[t]
\center{\includegraphics[width=12cm,height=9cm,angle=0]{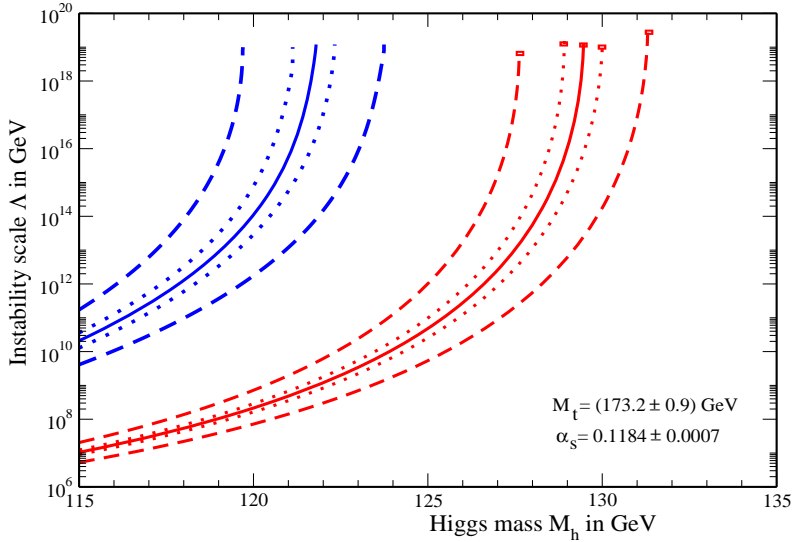}}
\caption{\label{inst}\em The  scale $\Lambda$ at which the SM Higgs potential becomes
negative as a function of the Higgs mass 
for the central value of $m_t$ and $\alpha_s$ (plain red), as well as for
$\pm 2\sigma$ variations of $m_t$ (dashed red) and  $\alpha_s$ (dotted red).
The blue lines on the left are the metastability bounds (plain blue: central values of $m_t$ and $\alpha_s$;
dashed blue: $\pm 2\sigma$ variations of $m_t$). The theoretical error in the determination of
the instability scale is not shown. } 
\label{LAMBDA}
\end{figure}

\begin{figure}[t]
$$\includegraphics[width=\textwidth]{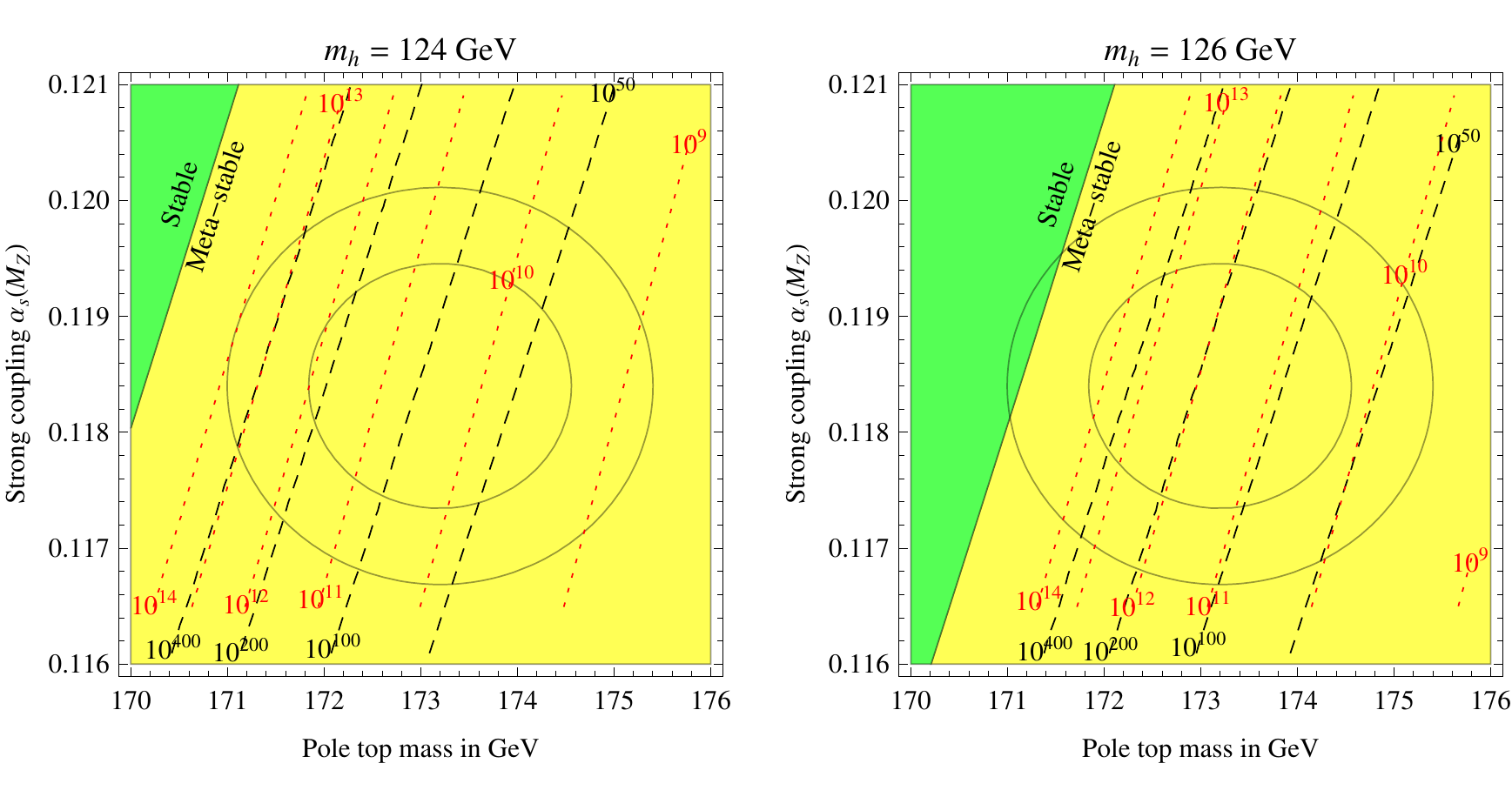}$$
\caption{\em Measured values of the top mass and of the strong coupling at 68, 95\% C.L. (2 dof) compared to the
regions of the parameter space which are stable (upper-left, shaded in green)
and meta-stable  (yellow).
In the latter case, the dashed curves are the iso-contours of the lifetime in years, and the dotted curves are the iso-contours
of the instability scale in $\GeV$.\label{fig:S} 
The theroretical error, estimated to be $\pm 3$~GeV in $m_h$ at fixed $m_t$,  
is not shown.}
\end{figure}

\begin{figure}[t]
$$\includegraphics[width=0.7\textwidth]{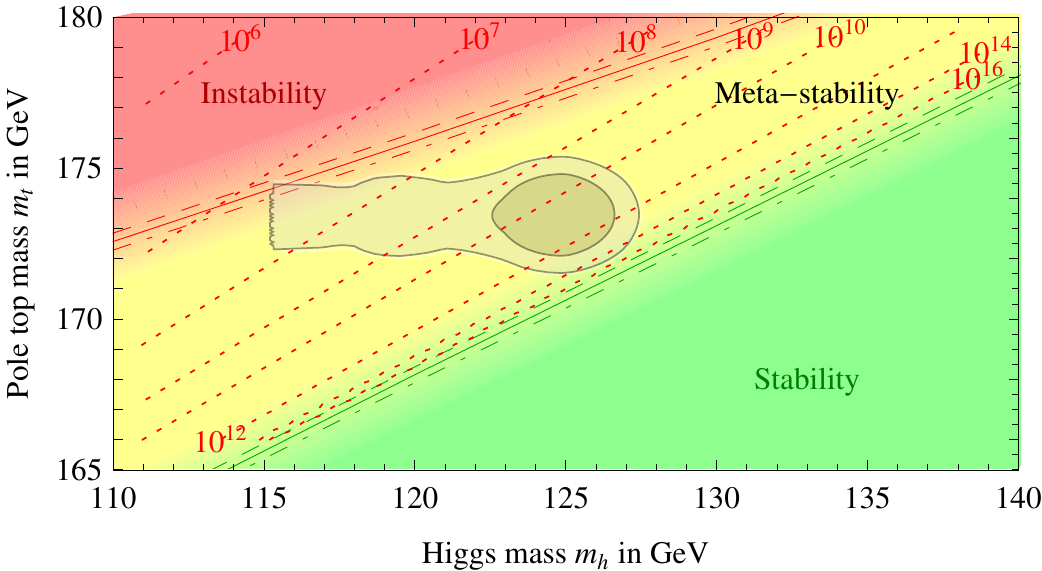}$$
\caption{\em Measured value of the top mass and preferred range of $m_h$, compared to the regions corresponding to
absolute stability, meta-stability and instability of the SM vacuum.
The three  boundaries lines corresponds to $\alpha_s(M_Z)=0.1184\pm 0.0007$, and the grading of the 
colors indicates the size of the theoretical errors.  
The dotted contour-lines show the instability scale $\Lambda$ in $\GeV$ assuming $\alpha_s(M_Z)=0.1184$.
\label{fig:regions}}
\end{figure}

\section{Stability and metastability bounds}
We first present the analysis on the Higgs instability region at zero temperature. 
We are concerned with   large
field field values and therefore it is adequate to neglect the
Higgs mass term and to approximate the potential of the real field $h$
contained in the Higgs doublet $H=(0,v+h/\sqrt{2})$ as
\beq
V=\lambda (|H|^2-v^2)^2  \approx
\frac{\lambda}{4}h^4\ . 
\label{V}
\eeq
Here $v=174$~GeV and the physical Higgs mass is $m_h=2v\sqrt{\lambda}$ at tree level.
Our study here follows previous state-of-the-art analyses (see in particular~\cite{CEQ,IRS,IRST}).
We assume negligible corrections to the Higgs 
effective potential from physics beyond the SM up to energy scales of the 
order of  the Planck mass. 
We include two-loop renormalization-group (RG) equations for all the SM couplings, and 
all the known finite one and two-loop corrections in the relations between $\lambda$ and the top Yukawa coupling ($y_t$)
to $m_t$ and $m_h$.\footnote{In particular, we include one-loop electroweak 
corrections in the determination of  $\lambda(m_t)$ and $y_t(m_t)$, as well as two-loop QCD 
corrections in the determination of $y_t(m_t)$. }
Working at this order, we obtain a  two-loop
renormalization-group improved Higgs potential. 
Similarly, the effective action relevant to vacuum decay is computed at next-to-leading order in all the relevant SM couplings, including the gravitational coupling~\cite{IRST}.
The main novelty of our analysis of the bounds is the use of the recent experimental information on the Higgs mass and of the updated 
values of the top mass and strong couplings:
\beq \label{mtalpha3} m_t = (173.2\pm0.9)\GeV~\hbox{\cite{topmass}},
\qquad \alpha_s (M_Z) = 0.1184\pm 0.0007~\hbox{\cite{alpha3}}\ .\eeq   

In fig.\fig{run} we show the quartic Higgs coupling renormalized at high scales in the $\overline{\rm MS}$ scheme.
We see that for $m_h=124~(126)\GeV$ and for the best-fit values of $m_t$ and of $\alpha_s$,
the coupling becomes negative around $10^9~(10^{10})\GeV$ (continuous line).
However, this scale can be shifted by orders of magnitude by varying $m_t$ within its uncertainty band:
the red (blue) dashed lines show the effect of increasing (decreasing) $m_t$ by two standard deviations.
In particular, reducing $m_t$ by two standard deviations allows to avoid the instability for $m_h=126$~GeV.
Furthermore,  the red (blue) dotted lines show the effect of increasing (decreasing) $\alpha_s$ by two standard deviations,
which has a smaller impact.

The instability scale $\Lambda$ of the Higgs potential,
defined as the Higgs vev at which the one-loop effective potential turns negative, is shown in fig.~\ref{inst}, (lowest red curves). 
This scale takes into account additional finite one-loop corrections \cite{CEQ} and is 
typically at least one order of magnitude above the scale at which $\lambda(\mu)=0$. 
The lines in fig.~\ref{inst} end at the scale at which the new minimum of the Higgs potential 
characterized  by a large vacuum expectation value becomes degenerate with the electroweak one.

Fig.\fig{S} shows the boundary between stability (green) and meta-stability (yellow) regions for fixed values of the Higgs mass,
and as a function of the top mass and of the strong coupling.
The condition for stability can be approximated as
\beq
m_h > 130\gev + 1.8 \gev \left( \frac{m_t -173.2 \gev}{0.9\gev} \right) 
-0.5\gev  \left( \frac{\alpha_s(M_Z)-0.1184}{0.0007}\right)~ \pm 3~\gev~,
\label{stability}
\eeq
where the error of 3~GeV is an estimate of unknown higher-order effects.\footnote{~The theoretical errors 
in eqs.~(\ref{stability}) and (\ref{quantum})
are dominated by the uncertainties in the determination of $\lambda(m_t)$ and 
$y_t(m_t)$ in terms of $m_h$, $m_t$ and the other SM couplings. In particular, the leading error is 
induced by the unknown two-loop finite corrections in the determination of $\lambda(m_t)$. 
Estimating the size of  these effects by varying the matching scale 
on $\lambda(\mu)$ in the range $m_t/2 < \mu < 2 m_t$ leads to 
$\pm 2$~GeV in  $m_h$. The $\pm 0.5$~GeV theoretical error~\cite{Hoang} in the relation
between the measured value of $m_t$  and $y_t(m_t)$ leads to an additional $\pm 1$~GeV.
Summing linearly these two errors leads to the final error in eqs.~(\ref{stability}) and (\ref{quantum}). }

Data indicate that we live close to this boundary, which corresponds to the 
intriguing possibility  of a vanishing Higgs coupling (and perhaps also its beta function) at the Planck scale --- a possibility discussed in previous papers with different motivations, 
see e.g.~\cite{IRST,Nima,Shap,Bardeen,Nielsen,HLL}. Note however, that the present experimental 
situation is only marginally compatible with the realization of such scenario. If there is indeed 
a potential instability below the Planck scale, the minimal scenario of Higgs inflation \cite{HiggsI}
(which already suffered from a unitarity/naturalness problem \cite{NoHiggsI}) cannot be realized and one would be lead to nonminimal options that should cure, not only the unitarity problem \cite{GLI} but also the instability (a potential threat to 
scenarios such as those proposed in ref.~\cite{MN}).

\smallskip

\subsection{Meta-stability}
The fact that the Higgs potential develops a new deeper minimum does not necessarily 
mean that the situation is inconsistent, because our Universe could live in a metastable  vacuum. 
As shown in fig.\fig{run}, the evolution of $\lambda$ for $124\GeV < m_h < 126\GeV$ is such that it never becomes
too negative, resulting in a very small probability of quantum tunneling. Updating our previous 
analyses, we find that the lifetime of the electroweak vacuum is longer than the age of the Universe for
\beq
m_h > 111 \gev +2.8\gev \left( \frac{m_t -173.2 \gev}{0.9\gev} \right) 
-0.9\gev  \left( \frac{\alpha_s(M_Z)-0.1184}{0.0007}\right) \pm 3\gev~.
\label{quantum}
\eeq

The regions of stability, metastability, and instability in the $m_h$--$m_t$ plane
are illustrated in fig.~\ref{fig:regions}. 
As can be seen, present data strongly favor metastability, although full stability 
is still allowed. In the metastable case, the expected lifetime of the SM vacuum, 
depending on the precise values of $m_t$ and $\alpha_s$,
for two representative values of $m_h$, is reported in fig.~\ref{fig:S}. 

These considerations clearly show the importance of future precise determinations of the top and Higgs masses that can be achieved at the LHC, together with higher-order theoretical computations.

\begin{figure}[t]
\center{
\includegraphics[width=12cm,height=9cm]{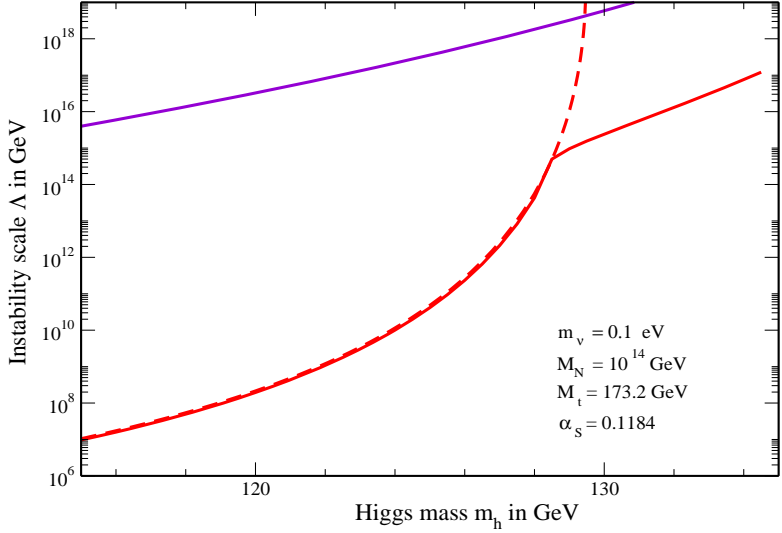}}
\caption{\label{fig:exnuRbound}\em Stability bound for $m_t=173.2$ GeV and $\alpha_s(M_Z)=0.1184$ with (solid red) and without (dashed red) right-handed neutrinos at
$M_N=10^{14}\GeV$ implementing a seesaw mechanism with $m_\nu=0.1\eV$. The purple line corresponds to the metastability bound against vacuum decay by quantum tunneling in the presence of such see-saw. (The corresponding limit without right-handed neutrino effects lies below the LEP limit on the Higgs mass.)}
\end{figure}

\subsection{Bounds on right-handed neutrinos}
\label{sect:RH}

\noindent
The presence of heavy right-handed neutrino states $N$ in addition to the 
SM particle content 
\beq \Lag = \Lag_{\rm SM} + i\bar N \slashed{\partial} N+
y_\nu LH N+\frac{M_N}{2} N^2+ \hbox{h.c.}\eeq
is well-motivated by the lightness of the left-handed neutrinos through
the see-saw mechanism,
\beq m_\nu = v^2 y_\nu \cdot M_N^{-1}\cdot y_\nu^T~.
\label{seesaw} 
\eeq 
Therefore it is interesting to analyze the impact of the RH neutrinos on the instability
region of the Higgs sector \cite{CdCIQ}.
In view of eq.~(\ref{seesaw}) the Yukawa couplings $y_\nu$ of the right-handed neutrinos
are sizable if right-handed neutrinos are heavy, and they
affect the RG evolution of the quartic higgs coupling making $\lambda$ more negative at large scales $\mu$ above $M_N$:
\beq (4\pi)^2 \frac{d\lambda}{d\ln\mu}=-2\Tr y_\nu y_\nu^\dagger y_\nu y_\nu^\dagger  -6 y_t^4 
+\frac{3}{8}\left[2 g^4 +(g^2+ {g'}^2)^2\right]+ {\cal O}(\lambda).
\label{eq:RGE}\eeq
We use the full two-loop RG equations,
and we assume three degenerate right-handed neutrinos at the mass $M_N$
with equal couplings $y_\nu$ which give three degenerate
left-handed neutrinos at mass $m_\nu = y_\nu v^2/M_N$. 
This is a plausible assumption as long as neutrino masses are larger that the observed
atmospheric neutrino mass difference $(\Delta m^2_{\rm atm})^{1/2} \approx 0.05\eV$.
Cosmological observations set an upper bound on $m_\nu$ of about 0.5 eV~\cite{nureview}.

\begin{figure}[t]
$$\includegraphics[width=0.5\textwidth]{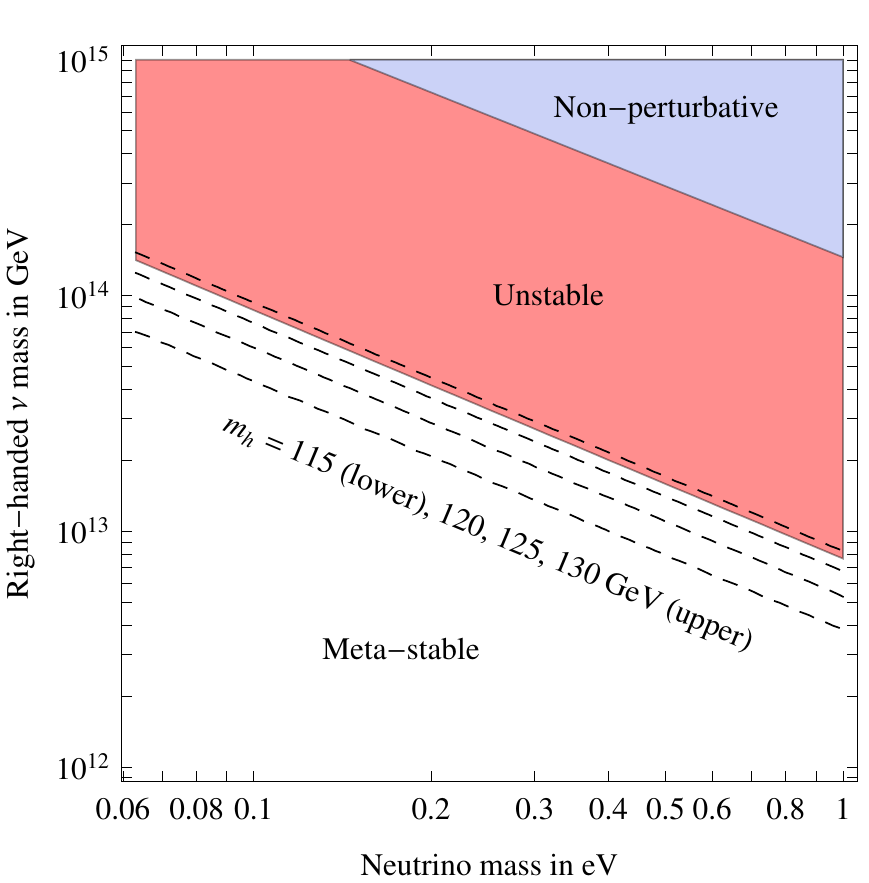}$$
\caption{\em Upper bound at 95\% C.L. on the right-handed neutrino mass from vacuum stability (region shaded in red)
as a function of the mass of light neutrinos, assumed to be degenerate.
The upper region shaded in gray corresponds to non-perturbative right-handed neutrino couplings, $y_\nu^2>6$.
\label{fig:nuRbound}}
\end{figure}

The effect of heavy right-handed neutrinos on the instability and metastability scales is illustrated
in  fig.~\ref{fig:exnuRbound} for one particular choice of parameters.
In fig.\fig{nuRbound} we show the upper bound on $M_N$ obtained imposing that the
lifetime of the SM vacuum exceeds the age of the universe. The bound 
ranges from $M_N < 10^{14}\GeV$ for $m_\nu \approx 0.05\eV$ to 
$M_N < 10^{13}\GeV$ for $m_\nu \approx 1\eV$, with a weak dependence 
on the Higgs mass within its favored range.
The stability bound roughly corresponds to $y_\nu <0.5$ and is therefore stronger than the perturbativity bound. 

We stress that this bound holds only if RH neutrinos are the only new degrees of freedom
appearing below the (high) energy scale where $\lambda$ turns negative. One can indeed consider also additional  degrees of freedom
that could contribute keeping $\lambda>0$. The best motivated case is, of course, 
supersymmetry, which ensures $\lambda>0$ by relating it to gauge couplings and, moreover, makes technically natural the hierarchy between the electroweak and see-saw scales \cite{susyseesaw}.
Two additional examples are 
i) extra quartic couplings of the form $S^2|H|^2$, where $S$ is a new light scalar which could be the Dark Matter particle, that
would contribute positively to \eq{eq:RGE};
ii) extra weak-multiplets around the electroweak scale (e.g.\ Dark Matter multiplets~\cite{MDM}), 
which increase the value of $g,g'$ at higher scales, indirectly
giving a positive contribution to \eq{eq:RGE}.

%
%
%
%
%
%
%
%

\section{Thermal vacuum decay}
\noindent
If the Universe underwent a period of expansion when the thermal plasma 
of relativistic degrees of freedom was at high temperature, 
thermal fluctuations  could have caused  the decay of the metastable electroweak vacuum
\cite{thermal,thermal1,thermal2} by nucleation of bubbles that probe the
Higgs instability region. On the other hand, high-temperature effects also
modify the Higgs effective potential, with a tendency of making the origin
more stable.

\begin{figure}[t]
\center{
\includegraphics[width=12cm,height=9cm]{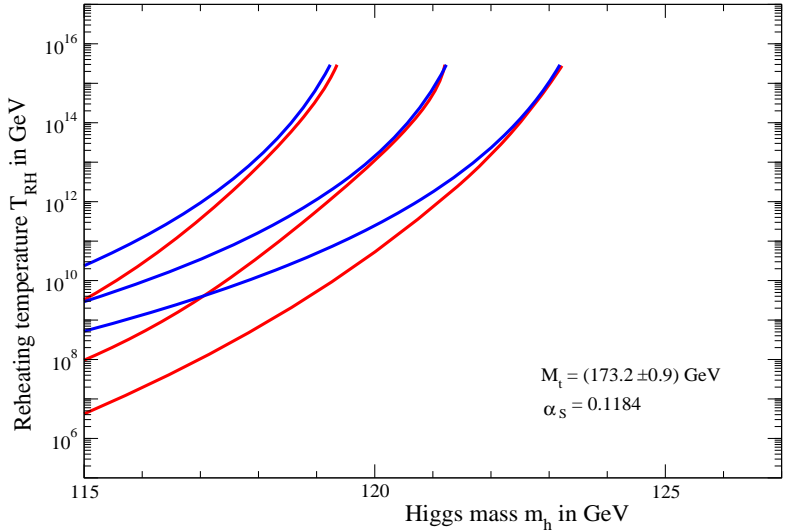}}
\caption{\label{fig:RHI}\em  Upper bounds on  the reheating temperature $T_{\rm RH}$, as functions of $m_h$, from sufficient 
stability of the electroweak vacuum against thermal fluctuations in the 
hot early Universe for three different values of the top mass (central value and $\pm 1 \sigma$).
The lower (red) curves are for $H_{\rm f}=10^{14}\GeV$, the upper ones for 
$H_{\rm f}=[4\pi^3 g_* (T_{\rm RH})/45]^{1/2}
(T_{\rm RH}^2/M_{\rm Pl})$, which corresponds to the case of instant reheating.
We take $\alpha_S(M_Z)=0.1184$. Lowering (increasing) 
$\alpha_s(M_Z)$ by one standard deviation lowers (increases) the bound on 
$T_{\rm RH}$ by up to one order of magnitude.} 
\label{mhTRH}
\end{figure}

Following the calculation in Ref.~\cite{us}, the requirement that the false vacuum does not decay 
during the high $T$ stages of the early Universe
sets an upper bound on the reheating temperature $T_{\rm RH}$ after inflation
once the Higgs mass is fixed. 
It should be remembered that $T_{\rm RH}$ in fact is not the 
maximal temperature achieved after inflation. Such maximal temperature 
occurs after inflation ends and before reheating completes and is given by~\cite{GKR}
\beq
\label{tmax}
T_{\rm max}=\left(\frac{3}{8}\right)^{2/5}
\left(\frac{5}{\pi^3}\right)^{1/8}
\frac{g^{1/8}_*(T_{\rm RH})}{g^{1/4}_*(T_{\rm max})}M^{1/4}_{\rm Pl}H_{\rm f}^{1/4}T_{\rm RH}^{1/2}
\ ,
\eeq
where $g_*(T)$ counts the effective number of degrees of freedom (with a 
$7/8$ prefactor for fermions) with masses $\ll T$ and $H_{\rm f}$ is 
the Hubble parameter at the end of 
inflation. The 
metastability bound on $T_{\rm RH}$ therefore depends on the particular value 
of $H_{\rm f}$: for a given $T_{\rm RH}$, the value of $T_{\rm max}$ grows with $H_{\rm f}$. 
Therefore the metastability constraint on $T_{\rm RH}$ will be more 
stringent for larger values of  $H_{\rm f}$ \cite{us}.

Figure~\ref{fig:RHI} shows the 
metastability bound on $T_{\rm RH}$ as a function of the 
 Higgs mass for various values of the top mass and for two choices of the
Hubble rate $H_{\rm f}$ at the end of inflation. 
The lower curves correspond to $H_{\rm f}=10^{14}$  GeV while the upper ones 
have 
\beq H_{\rm f}= H_{\rm f}^{\rm min}\equiv [4\pi^3 g_* (T_{\rm RH})/45]^{1/2}
(T_{\rm RH}^2/M_{\rm Pl})\eeq 
which is the lowest value of $H_{\rm f}$ allowed once it is required that 
 the inflaton energy density 
$\rho_\phi =3M_{\rm Pl}^2 H_{\rm f}^2/(8\pi)$ is larger than the energy density of a 
thermal bath with temperature $T_{\rm RH}$.
The current observations of the Cosmic Microwave Background (CMB) anisotropies
 \cite{wmap} are consistent with a smooth 
and nearly Gaussian power spectrum of curvature perturbations limiting the contributions to the anisotropies from 
of tensor modes. This translates into an upper bound of the Hubble rate during inflation
given by $H_{*}<4\times 10^{14}$ GeV. Since the Hubble rate during inflation decreases, that is $H_{\rm f}<H_*$, 
the corresponding maximal upper bound on $T_{\rm RH}$ is 
 $T_{\rm RH}< 2.6\,[106.75/g_*(T_{\rm RH})]^{1/2}\times 10^{15}$ GeV.

The bound on $T_{\rm RH}$ from thermal metastability gets weaker for smaller
values of the top mass or larger values of the 
Higgs mass since the instability scale becomes higher. 
Figure~\ref{fig:RHI}
shows the thermal metastability scale (blue highest lines), defined as the scale 
below which the Higgs potential should be modified to avoid thermal decay.
For
\beq
m_h > 121.7\gev + 2 \gev \left( \frac{m_t -173.2 \gev}{0.9\gev} \right) 
-0.6\gev  \left( \frac{\alpha_s(M_Z)-0.1184}{0.0007}\right) \pm 3\gev \, ,
\label{metastability}
\eeq
the vacuum is sufficiently long-lived even for $T\sim M_{\rm Pl}$,
and therefore, the bound on $T_{\rm RH}$ disappears~\cite{thermal2}.  
This is the reason why the lines in fig.~\ref{fig:RHI} stop at some value 
of $m_h$. 
From this figure it is also clear that the bound on $T_{\rm RH}$ is 
very sensitive on the value of $m_t$, with the experimental error in $m_t$ 
being the main source of uncertainty.

\begin{figure}
\center{
\includegraphics[width=12cm,height=9cm]{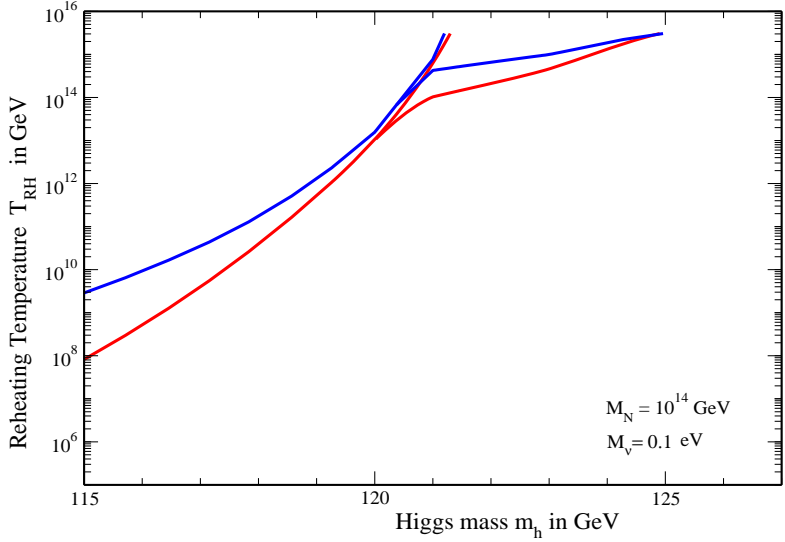}}
\caption{\label{fig:RH2}\em 
Upper bounds on  the reheating temperature $T_{\rm RH}$, as functions of $m_h$, 
from sufficient  stability of the electroweak vacuum against thermal fluctuations in the 
hot early Universe, as in fig.~\ref{fig:RHI}, but with a seesaw mechanism with $M_N=10^{14}~\gev$, $m_\nu=0.1\eV$. As in fig.~\ref{fig:RHI}, the lower curves are 
for $H_{\rm f}=10^{14}~\gev$, the upper ones for instant reheating, and for the central values of  $m_t$ and $\alpha_s$. }
\end{figure}

\subsection{Thermal vacuum decay and leptogenesis}

Let us briefly discuss  what are the implications of the bound on $T_{\rm RH}$ shown in 
fig.~\ref{fig:RHI} for the dynamics of the evolution of the Universe.
As we mentioned in the introduction, an appealing  mechanism for the 
generation of the baryon asymmetry is leptogenesis
in which the asymmetry is produced by the out-of-equilibrium decay of heavy RH neutrinos. Of course, the mechanism
works  as long as the Universe was ever populated by such heavy states during its evolution.
In the most popular version of leptogenesis, the so-called thermal leptogenesis, these heavy
states are produced through thermal scatterings. This  sets a lower bound on $T_{\rm RH}$ as a function of $M_1$, the mass
of the lightest right-handed neutrino~\cite{DI}. This bound reaches
its minimum for $M_1 \sim T_{\rm RH}$, when $T_{\rm RH}>3\times
10^9\gev$~\cite{thlepto}.

The first conclusion we can draw is that hierarchical thermal leptogenesis  is not allowed  if the Higgs mass turns out
to be less than about 120 GeV and the top mass is on the high
side of the allowed experimental range. 

Moreover, we can conclude that the value of $m_h=125$ GeV is  consistent 
with thermal leptogenesis and a hierarchical spectrum of RH neutrinos. In fact, for such a value of the Higgs mass, the reheating temperature is  comfortably  high so that   any baryogenesis mechanism would be operative.

As discussed in section~\ref{sect:RH}, the Yukawa couplings $y_\nu$ of the heavy
right-handed neutrinos can also modify the instability scale of the
Higgs potential~\cite{CdCIQ} and this, in turn, affects the bound on $T_{\rm RH}$.
These effects turn out to be important only if the mass of the right-handed neutrinos is
sufficiently large, $M_1\simgt (10^{13}-10^{14})\gev$ ~\cite{CdCIQ}. The effect is
illustrated in fig.~\ref{fig:RH2}, which shows the impact of RH neutrinos with $M_N=10^{14}$~GeV 
(with $m_\nu=0.1$~eV) on the bound on the reheating temperature.
We conclude that the existence of heavy right-handed neutrinos affect the bounds on $T_{\rm RH}$ 
for a larger interval of Higgs masses, but only for $T_{\rm RH}>M_1\simgt \left(10^{13}-10^{14}\right)\gev$.

We stress that these considerations apply only to the case of hierarchical thermal leptogenesis
in the SM, with no new physics present below the scale $M_1$. Thermal leptogenesis 
may indeed occur with almost degenerate RH neutrinos, allowing much lighter 
values for $M_1$ (as low as the TeV scale). In such case, no relevant constraint can be derived from stability considerations. 

\section{Conclusions}
We have analysed the stability of the Standard Model vacuum with special emphasis on the 
hypothesis that the Higgs mass, $m_h$, is in the following range:
$124~{\rm GeV} < m_h <126$~GeV, as hinted by recent ATLAS and CMS data.

Given the upper bound on the Higgs mass of 127 GeV, we conclude that 
the Standard Model ground state is very likely to be metastable. 
In  the preferred range of $m_h$ the  deeper minimum of the potential 
occurring at very high energies is sufficiently long-lived compared 
to the age of the Universe. Full stability is unlikely and would require 
$m_t$ to be closer to its lower allowed range, as summarized in fig.~\ref{fig:S}.

The scale where the Higgs potential becomes unstable is very high, 
around $10^{11}$~GeV for $m_h=125$~GeV and central values of
$m_t$ ans $\alpha_s$.  As a result,  no significant constrains on the 
reheating temperature are obtained. On the other hand, if the model 
is extended with the inclusion of heavy RH neutrinos, an upper bound
on their masses in the $10^{13}$--$10^{14}\GeV$ range
(summarized in fig.~\ref{fig:nuRbound}) can be derived by the 
requirement that the electroweak vacuum 
has a lifetime longer than the age of the Universe.

\small

\paragraph{Acknowledgments}
We thank A. Hoecker for discussions on the uncertainty in $\alpha_s$.
This work was supported by the ESF grant MTT8 and by SF0690030s09 project.
J.R.E. thanks G. Servant for requesting updates of the results in \cite{us}.
Work supported in part by  the Spanish 
Ministry MICINN under contracts FPA2010-17747 and FPA2008-01430; the Spanish Consolider-Ingenio 2010 Programme CPAN (CSD2007-00042); and the Generalitat de Catalunya grant 2009SGR894. 
J. Elias-Mir\'o is supported by an FPU grant from the Spanish Ministry of Science and Innovation.
G.I. acknowledges the support of the Technische Universit\"at M\"unchen -- Institute for Advanced
Study, funded by the German Excellence Initiative, and MIUR under contract 2008XM9HLM.

\small
\begin{multicols}{2}
\end{multicols}
\end{document}